\documentclass[twocolumn]{aastex701}

\usepackage{xcolor}

\usepackage{booktabs}
\usepackage{array}
\usepackage{amsmath}
\usepackage{multirow}
\usepackage{enumitem}

\begin{document}

\title{Finding Quasars behind the Galactic Plane. IV. Candidate Selection from \emph{Chandra} with Random Forest}

\author[gname=Xu, sname=Zhang]{Xu Zhang} 
\affiliation{Shenzhen Key Laboratory of Ultraintense Laser and Advanced Material Technology, Center for Intense Laser Application Technology, and College of Engineering Physics, Shenzhen Technology University, Shenzhen 518118, People's Republic of China}
\email{2410262039@stumail.sztu.edu.cn}  

\author[orcid=0000-0003-4897-4106]{Yanli Ai}
\affiliation{Shenzhen Key Laboratory of Ultraintense Laser and Advanced Material Technology, Center for Intense Laser Application Technology, and College of Engineering Physics, Shenzhen Technology University, Shenzhen 518118, People's Republic of China}
\email[show]{aiyanli@sztu.edu.cn}

\author[orcid=0000-0002-6610-5265]{Yanxia Zhang} 
\affiliation{National Astronomical Observatories, Beijing 100101, People's Republic of China}
\email{zyx@bao.ac.cn}

\author[orcid=0000-0002-0759-0504]{Yuming Fu} 
\affiliation{Leiden Observatory, Leiden University, Einsteinweg 55, 2333 CC Leiden, The Netherlands}
\affiliation{Kapteyn Astronomical Institute, University of Groningen, PO Box 800, 9700 AV Groningen, The Netherlands}
\email{yfu@strw.leidenuniv.nl}

\author[orcid=0000-0002-7350-6913]{Xue-Bing Wu} 
\affiliation{Department of Astronomy, School of Physics, Peking University, Beijing 100871, People's Republic of China}
\affiliation{Kavli Institute for Astronomy and Astrophysics, Peking University, Beijing 100871, People's Republic of China}
\email{wuxb@pku.edu.cn}

\author[orcid=0009-0003-3066-2830]{Zhiying Huo}
\affiliation{National Astronomical Observatories, Beijing 100101, People's Republic of China}
\email{zhiyinghuo@bao.ac.cn}

\author{Wenfeng Wen} 
\affiliation{Shenzhen Key Laboratory of Ultraintense Laser and Advanced Material Technology, Center for Intense Laser Application Technology, and College of Engineering Physics, Shenzhen Technology University, Shenzhen 518118, People's Republic of China}
\email{2300411002@stumail.sztu.edu.cn}

\author{Jiayuan Zhou} 
\affiliation{Shenzhen Key Laboratory of Ultraintense Laser and Advanced Material Technology, Center for Intense Laser Application Technology, and College of Engineering Physics, Shenzhen Technology University, Shenzhen 518118, People's Republic of China}
\email{202200802018@stumail.sztu.edu.cn}

\author{Dexuan Kong} 
\affiliation{Shenzhen Key Laboratory of Ultraintense Laser and Advanced Material Technology, Center for Intense Laser Application Technology, and College of Engineering Physics, Shenzhen Technology University, Shenzhen 518118, People's Republic of China}
\email{2510262102@stumail.sztu.edu.cn}

\author{Linfeng Zeng} 
\affiliation{Shenzhen Key Laboratory of Ultraintense Laser and Advanced Material Technology, Center for Intense Laser Application Technology, and College of Engineering Physics, Shenzhen Technology University, Shenzhen 518118, People's Republic of China}
\email{202300802061@stumail.sztu.edu.cn}

\author{Heng Wang} 
\affiliation{Shenzhen Key Laboratory of Ultraintense Laser and Advanced Material Technology, Center for Intense Laser Application Technology, and College of Engineering Physics, Shenzhen Technology University, Shenzhen 518118, People's Republic of China}
\email{aiyanli@sztu.edu.cn}

\correspondingauthor{Yanli Ai}  

\begin{abstract}
Quasar samples remain severely incomplete at low Galactic latitudes because of strong extinction and source confusion. We conduct a systematic search for quasars behind the Galactic plane using X-ray sources from the \emph{Chandra} Source Catalog (CSC~2.1), combined with optical data from \textit{Gaia} DR3 and mid-infrared data from CatWISE2020. Using spectroscopically confirmed quasars and stellar-type objects from data sets including DESI, SDSS, and LAMOST, we apply a Random Forest classifier to identify quasar candidates, with stellar contaminants suppressed using \textit{Gaia} proper-motion constraints. Photometric redshifts are estimated for the candidates using a Random Forest regression model. Applying this framework to previously unclassified CSC sources, we identify 7570 quasar candidates, including 1060 Galactic Plane Quasar (GPQ) candidates at $|b|<20^{\circ}$, of which 551 are high-confidence candidates. Relative to the previously known GPQ sample, our selected GPQs reach lower optical and X-ray fluxes, improving sensitivity to low-flux GPQs. In addition, both the GPQ candidates and known GPQs display harder X-ray spectra than the all-sky quasar sample, consistent with increased absorption through the Galactic plane. Pilot spectroscopy confirms two high-confidence GPQ candidates as quasars at spectroscopic redshifts of $z=1.2582$ and $z=1.1313$, and further spectroscopic follow-up of the GPQ sample is underway. This work substantially improves the census of GPQs and provides a valuable target sample for future spectroscopic follow-up, enabling the use of GPQs to refine the reference frames for astrometry and probe the Milky Way interstellar and circumgalactic media with the absorption features of GPQs.

\end{abstract}

\section{INTRODUCTION} 

Quasars are among the most luminous active galactic nuclei (AGNs), powered by accretion onto supermassive black holes, and are detectable out to cosmological distances. Their high luminosities, compactness, and long-lived activity phases make quasars uniquely valuable probes for a wide range of astrophysical and cosmological questions, including the growth of supermassive black holes and their coevolution with host galaxies (e.g., \citealt{DiMatteo2005, Kormendy_2013, Wu_2015, Banados_2018, Inayoshi_2020, Fan_2023}), and the large-scale structure of the Universe through quasar clustering and baryon acoustic oscillations (e.g., \citealt{Eisenstein_2005, Dawson2013, Blanton2017}). In addition, quasars have negligible parallaxes and proper motions, making them ideal reference objects for defining and maintaining celestial reference frames (CRFs; e.g., \citealt{Ma2009, Gaia_early_2016, Gaia_2023a}). Despite major progress in wide-area surveys, however, the completeness of quasar samples remains poor close to the Galactic plane.

Over the past decades, major spectroscopic surveys have dramatically expanded the known quasar population. The Sloan Digital Sky Survey \citep[SDSS;][]{SDSS, York2000} and its extensions established the first homogeneous, large-area quasar samples \citep[][]{Richards2002, Schneider2010, Paris2017, Lyke2020}, while complementary efforts such as the 2dF QSO Redshift Survey and the LAMOST quasar survey extended quasar identification to different depths and sky regions \citep{Croom2004, Croom2009, Ai2016, Jin2023, Lyu2025}. More recently, the Dark Energy Spectroscopic Instrument \citep[DESI;][]{DESI_DR1} has delivered the largest spectroscopic quasar samples to date, with more than one million confirmed quasars now available for statistical studies. Despite this rapid progress, the sky distribution of known quasars remains highly non-uniform.

In particular, the census of quasars at low Galactic latitudes remains severely incomplete. Most optical and near-infrared surveys avoid regions close to the Galactic plane, where strong interstellar extinction, high stellar densities, and severe source confusion substantially degrade the effectiveness of traditional color-based selection techniques \citep{Schlegel1998, Ross2012, Richards2009}. As a consequence, the surface density of spectroscopically confirmed quasars decreases sharply toward the Galactic plane, producing a persistent gap in all-sky quasar samples \citep{Fu_2021, Huo2025}. This incompleteness limits the use of quasars as uniform tracers of large-scale structure and as astrometric reference objects, and motivates dedicated searches for quasars in the low-latitude sky.

Quasars behind the Galactic plane (GPQs; $|b| \lesssim 20^\circ$), are nevertheless of considerable scientific importance. As distant extragalactic point sources with negligible parallaxes and proper motions, GPQs provide stable reference objects that are essential for constructing and validating celestial reference frames in the Galactic disk, where background sources are currently sparse (\citealt{Mignard2018, GaiaCollaboration2018, GaiaCollaboration2023}). 
GPQs also serve as bright background beacons for absorption-line studies of the Milky Way interstellar and circumgalactic media, enabling measurements of gas kinematics, dust extinction, and chemical abundances along heavily obscured sightlines that are under-sampled at high Galactic latitudes (e.g., \citealt{Savage1996, Richter2006, Fu_2021, Fu_2024}).

Identifying GPQs is observationally challenging. The Galactic plane hosts a dense and diverse population of stars, binaries, and other Galactic sources, including many X-ray emitters, and these contaminants can overlap quasars in color--color and color--magnitude spaces. As a result, GPQs have been explored in relatively few dedicated studies. Early efforts by \citet{Im2007} uncovered dozens of bright QSOs/AGNs at $|b| < 20^\circ$ using radio and near-infrared criteria. A major advance was achieved by \citet{Fu_2021}, who applied transfer learning to optical and infrared data from Pan-STARRS1 \citep[PS1;][]{PS1} DR1 and AllWISE \citep{Wright2010, Cutri2013}, producing a catalog of roughly 160,000 GPQ candidates within $|b| < 20^\circ$. Subsequent follow-up spectroscopy confirmed approximately 200 GPQs with high success rates \citep{Fu_2022}. \citet{Werk2024} later confirmed 72 UV-bright AGNs at $|b| < 30^\circ$, with 25 newly recognized objects, and the LAMOST GPQ program has recently reported $\sim 1300$ additional GPQs based on LAMOST DR10 \citep{Huo2025}. These studies demonstrate that combining multiwavelength data with modern classification approaches is essential for making progress in the Galactic plane.

Machine-learning methods applied to multi-band photometry and astrometry can exploit high-dimensional information to improve quasar selection in crowded fields \citep[][]{Richards2009, Carrasco2015, Gaia_2023a, Fu_2025}. X-ray observations provide a powerful and complementary avenue for identifying quasars with extreme properties \citep[][]{Wang2025} or X-ray sources in the Galactic disk \citep[][]{Bao2025}. Among current X-ray facilities, the \emph{Chandra} X-ray Observatory is uniquely advantageous for low-latitude work: its sub-arcsecond angular resolution substantially mitigates source confusion in dense stellar regions, enabling more reliable associations between X-ray sources and their optical/infrared counterparts. The \emph{Chandra} Source Catalog (CSC; here we use CSC~2.1) therefore offers an excellent basis for constructing GPQ candidates, provided that Galactic X-ray source contamination can be effectively controlled.

In this work, we present the fourth paper in the ``Finding Quasars behind the Galactic Plane'' series, focusing on GPQ candidate selection using CSC 2.1 and a Random Forest (RF) classifier. We cross-match CSC sources with \textit{Gaia} DR3 and CatWISE2020 to assemble optical, mid-infrared, and X-ray features for classification. To reduce contamination from extended sources and problematic photometry in crowded regions, we apply quality screening based on the \textit{Gaia} corrected BP/RP flux excess factor ($C^\ast$). We build a spectroscopically confirmed training set by cross-matching with DESI, SDSS, and LAMOST catalogs, train an RF classifier to identify quasar candidates, and then incorporate \textit{Gaia} proper motions to further suppress stellar contaminants. We also estimate photometric redshifts for the quasar candidates using an RF regressor. The resulting catalog contains 6286 quasar candidates in total, including 863 GPQ candidates, with a substantial high-confidence subset. Finally, we report initial spectroscopic confirmation of two GPQ candidates, as part of an ongoing follow-up program.

This paper is organized as follows. In Section \ref{sec:data}, we describe the CSC~2.1 data and the multi-wavelength photometric and astrometric information used in this work, as well as the construction of the training sample. Section \ref{sec:method} introduces the RF classification framework, the adopted features, and the performance of the classifier. In Section \ref{sec:redshift}, we present the RF-based photometric redshift estimation for the quasar candidates. The classification results and the properties of the selected GPQ candidates are discussed in Section \ref{sec:result}. Finally, we summarize our main results and conclusions in Section \ref{sec:summary}. We adopt a flat universe with $\Omega_{\rm M} = 0.3$ and $\Omega_{\Lambda} = 0.7$, and a cosmological parameter $H_0 = 70$ $\rm km$ $\rm s^{-1}$ $\rm Mpc^{-1}$.

\section{Data} \label{sec:data}

\subsection{CSC2.1 and Multi-Wavelength Data} 
\label{subsec:2.1}
We used \textit{Chandra} Source Catalog version 2.1 \citep[CSC~2.1;][]{CSC} Master Sources Table to select GPQ candidates. It includes measured properties for 407,806 unique compact and extended X-ray sources and more than 1.3 million individual detections observed with either ACIS or HRC-I observations released publicly before the end of 2021. Most of these sources remain unidentified. \textit{Chandra} has two focal plane instruments: Advanced CCD Imaging Spectrometer (ACIS) and High Resolution Camera (HRC). The ACIS instrument observes in broad (b): 0.5–7.0 keV, ultrasoft (u): 0.2–0.5 keV, soft (s): 0.5–1.2 keV, medium (m): 1.2–2.0 keV, and hard(h): 2.0–7.0 keV bands. The HRC instrument observes in 0.1–10 keV energy band and is designated as ‘W’ band. The energy flux in each band is determined using aperture photometry. The source count is derived from an elliptical source region and subtracted by the background count in the surrounding region. To convert the count rate to energy flux, the total count rate is summed up and then scaled by the local ancillary response function. 

In this work, we use aperture-corrected average net-flux in b, u, s, m, and h bands, which are named as $b$-csc, $u$-csc, $s$-csc, $m$-csc, $h$-csc, respectively. We applied several constraints to the CSC 2.1 data to ensure quality. First, we selected only point sources with the extent flag set to 0 and a detection significance of likelihood $> 10$. We then applied additional filters based on the following CSC 2.1 quality flags: pileup\_flag, sat\_src\_flag, conf\_flag, streak\_src\_flag (Table \ref{tab:quality_flag}). Through this process, a total of 357,727 unique high-quality X-ray point sources were selected from CSC 2.1.

\begin{table}[t]
\caption{Quality Flags Used to Filter Sources in CSC 2.1 and Their Description}
  \raggedright 
  \setlength{\tabcolsep}{2pt}
  \raggedright
  \begin{tabular*}{\columnwidth}{ll}
    \hline\hline
    \textbf{Flag code} & \textbf{Description} \\
    \hline
    extent\_flag & Extended or non-point-like at 90\% CL. \\
    pileup\_flag & ACIS pile-up fraction exceeds $\sim$10\% \\
    sat\_src\_flag & Saturated source in all observations \\
    conf\_flag & Source confused (source and/or background \\
    &regions in different stacks may overlap) \\
    streak\_src\_flag & Source on ACIS CCD readout streak\\
    \hline
  \end{tabular*}
  \label{tab:quality_flag}
\end{table}

We cross-match these 357,727 unique CSC~2.1 sources with the CatWISE2020 \citep{CatWISE} and \textit{Gaia} DR3 catalogues to obtain their multi-wavelength properties. Using the TOPCAT software \citep{TOPCAT}, we adopt matching radii of $3\arcsec$ and $1\farcs5$ for CatWISE2020 and \textit{Gaia} DR3, respectively, retaining only the nearest counterpart.

For the CatWISE2020 counterparts, we apply filters to retain sources with magnitudes in the ranges $8 < W1 < 17.7$ and $7 < W2 < 17.5$, and corresponding magnitude errors $\sigma \leq 0.2$, ensuring that the magnitudes are below the nominal limiting magnitudes in each band. For the \textit{Gaia} DR3 counterparts, we require $G < 21$~mag and a photometric error less than 0.2. 

All \textit{Gaia} magnitudes are then corrected for extinction using the two-dimensional dust reddening map from \citet{Planck} and the \citet{Fitzpatrick1999} extinction law assuming a total-to-selective extinction ratio of $R_V = 3.1$. To mitigate the potential impact of bright sources on classification performance, we further excluded objects with extinction-corrected $G$-band magnitudes brighter than 10.

To improve the purity of the GPQ candidate sample, we then remove potential optically extended sources using the following procedure. We refine the point-source selection by applying the corrected BP/RP photometric excess factor, $C^{*}$, derived from \textit{Gaia} DR3 \citep[][]{Gaia_cut}, which provides an additional constraint on source morphology. The original BP/RP excess factor, $C$, is defined as the ratio of the summed BP and RP integrated fluxes to the $G$-band flux:
\begin{equation}
C = \frac{I_{\mathrm{BP}} + I_{\mathrm{RP}}}{I_{G}} .
\end{equation}
Because the BP and RP photometric windows are broader than that of the $G$ band, extended sources tend to exhibit systematically larger excess factors than point sources \citep[e.g.,][]{Liu2020}. The corrected excess factor $C^{*}$ mitigates the color dependence inherent in $C$, providing a more reliable diagnostic of source extension \citep[][]{Gaia_cut}. To compress the dynamic range and suppress the influence of extreme outliers, we impose the criterion $\log(1 + C^{*}) \leq 0.5$, which yields a cleaner point-source sample \citep[e.g.,][]{Fu_2025}. Finally, 36,158 sources are selected with high-quality measurements from CSC~2.1, CatWISE2020, and \textit{Gaia} DR3, among which 15,588 are located within the Galactic plane ($|b| < 20^\circ$). Hereafter, we refer to this sample as the CSC--WISE--GAIA sample.

During the cross-matching procedure, a single X-ray source may be associated with multiple optical/infrared counterparts, leading to potential source confusion. This effect is particularly pronounced in the Galactic plane due to high extinction and stellar density. Indeed, among the 36,158 X-ray sources in our CSC--WISE--GAIA sample, 2,269 ($\sim$6\%) possess multiple \textit{Gaia} or CatWISE2020 counterparts within the matching radii. As expected, this confusion rate rises to $\sim$10\% (1,527 out of 15,588 sources) for the subset located within the Galactic plane ($|b| < 20^\circ$).

To evaluate the impact of source confusion, we conducted a parallel test using the probabilistic algorithm \texttt{nway} \citep{nway}. We found that while it improves sample purity, it does so at the expense of completeness (detailed in Section \ref{subsec:5.1}). Because our primary objective is to retain a larger sample of quasar candidates in the Galactic plane, we adopted a nearest-neighbor matching strategy.

\subsection{The Training Set}\label{subsec:2.2}

A well-defined and representative training set is critical for the supervised classification of quasars and stellar-type objects, particularly in the crowded and heterogeneous environments near the Galactic plane. Following the construction of the CSC--WISE--GAIA sample described in Section \ref{subsec:2.1}, we compile spectroscopically confirmed quasar and stellar samples from multiple surveys and catalogs to serve as labeled data for training the RF classifier.

\begin{figure*}[ht]
\centering
\includegraphics[width=\textwidth]{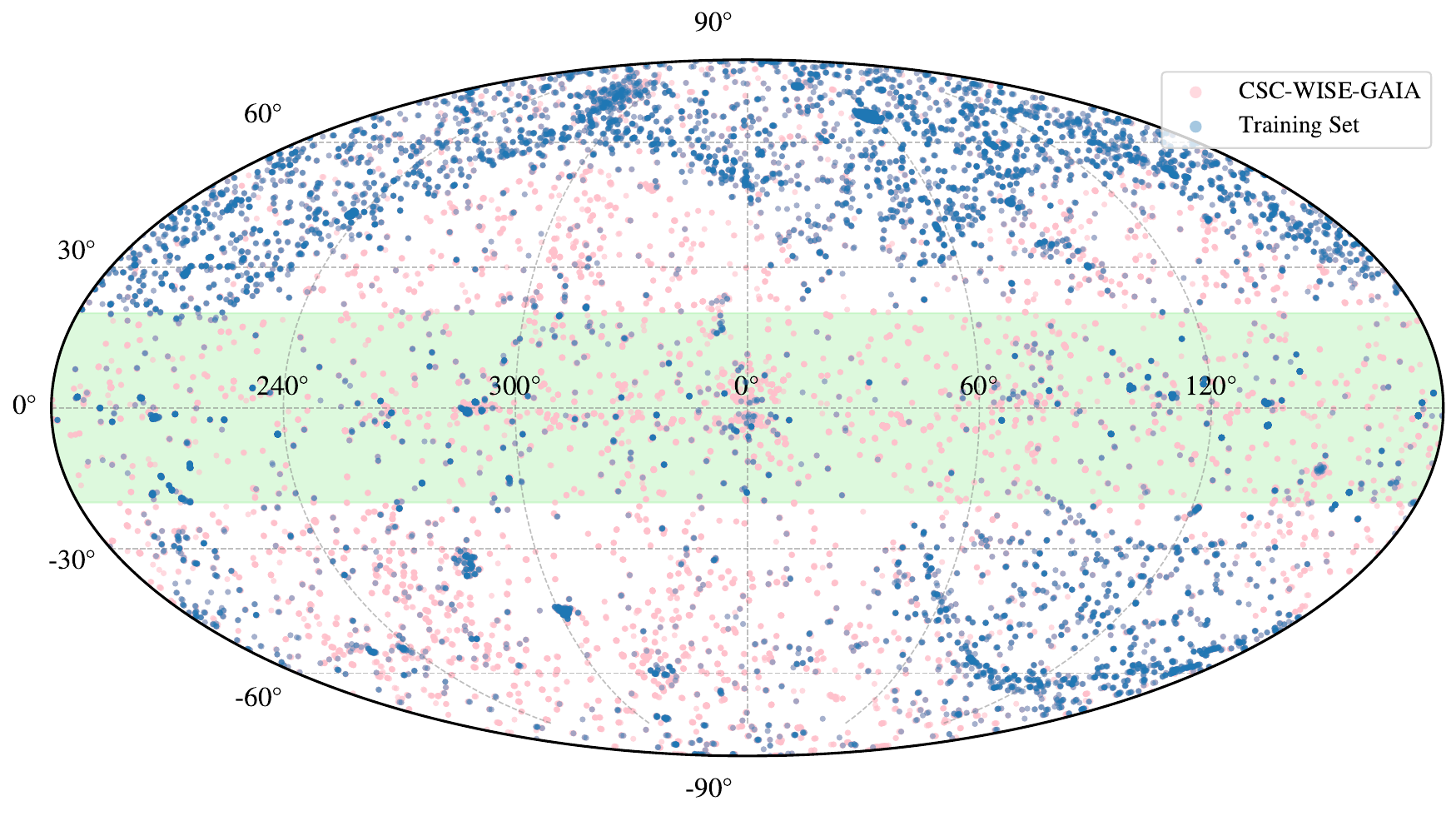}
\caption{Spatial distribution of the CSC-WISE-GAIA sample ($N=36,158$, pink) and the training set ($N=13,724$, blue) in Galactic coordinates using a Mollweide projection. The light green shaded region indicates the Galactic plane ($|b|<20^{\circ}$). This distribution demonstrates that while the training labels are primarily concentrated at high latitudes, our parent sample provides comprehensive coverage of the Galactic plane, enabling the discovery of new candidates in obscured regions.}
\label{fig:spatial_dist}
\end{figure*}

We first identify quasars by cross-matching the CSC--WISE--GAIA sample with spectroscopically confirmed quasars from DESI DR1 \citep{DESI_DR1} and SDSS DR18 \citep{SDSS}, adopting a matching radius of $1\arcsec$. To improve completeness, this initial quasar sample is further supplemented with confirmed quasars from the Million Quasars Catalog \citep{Milliquas}, the LAMOST spectroscopic surveys \citep{LAMOST_01,LAMOST_02}, and the Galactic Plane quasar sample identified from LAMOST DR10 by \citet{Huo2025}. All supplementary quasar catalogs are cross-matched with the CSC--WISE--GAIA sample using a $3\arcsec$ radius to ensure consistent source associations.

For the stellar class, we similarly cross-match the CSC--WISE--GAIA sample with DESI DR1 and SDSS DR18 to identify spectroscopically classified stars. To better represent the diverse Galactic X-ray source population that may contaminate quasar selection, we further include several specialized stellar catalogs. These include Galactic Wolf--Rayet stars \citep{derHucht2001,derHucht2006} and APOGEE-2 stars from SDSS DR16 \citep{Jonsson2020}; young stellar objects (YSOs) in open clusters and molecular clouds \citep[e.g.,][]{Ozawa2005,Giardino2007,Delgado2011,Povich2011,Rebull2011,Megeath2012}; high-mass X-ray binaries (HMXBs) and related Galactic and nearby extragalactic sources \citep{Liu2006,Walter2015,Mineo2012,Sazonov2017}; low-mass X-ray binaries (LMXBs) compiled from the Ritter--Kolb catalog and subsequent updates including recent comprehensive catalogs \citep{Ritter2015,Liu2007,Sazonov2020,Kundu2007,Humphrey2008,Zhang2011,Avakyan2023,Neumann2023}; and cataclysmic variables (CVs) \citep{Downes2005,Ritter2015}. These catalogs are cross-matched with the CSC--WISE--GAIA sample using a matching radius of $3\arcsec$. All sources matched to DESI/SDSS stars or identified in the aforementioned stellar catalogs are classified as stellar-type objects in this paper.

\begin{table}[t]
  \caption{The Number of Sources in the Training Set}
  \label{tab:number_of_source}
  \begin{tabular*}{\columnwidth}{@{\extracolsep{\fill}} lcr @{}}
    \hline\hline
    \textbf{Class} & \textbf{\% of training set} & \textbf{No.} \\
    \midrule
    QSO            & 61.86                        & 8489            \\
    STAR           & 38.14                        & 5235            \\
    \textbf{Total Training Set} & & \textbf{13,724}  \\
    Unidentified Sources        & & 22,434           \\
    \textbf{Total No.}          & & \textbf{36,158}  \\
    \hline 
  \end{tabular*}
    \par
    \vspace{1ex} 
    \footnotesize 
    \textbf{Note.} ``QSO'' and ``STAR'' represent quasars and stellar-type objects, respectively.
\end{table}

Sources that are not matched to any of the above quasar or stellar catalogs are subsequently queried against the SIMBAD database \citep{Wenger2000} within a $3\arcsec$ radius. Entries flagged as ``NOT GOOD'' are excluded. After applying these procedures, the final training set consists of 8489 quasars and 5235 stellar-type objects, as summarized in Table~\ref{tab:number_of_source}. The remaining 22,434 sources in the CSC--WISE--GAIA sample lack reliable spectroscopic classifications and are treated as unlabeled objects for subsequent prediction.

The spatial distribution of the final datasets is illustrated in Figure \ref{fig:spatial_dist}. We present the $N=36,158$ sources in the parent CSC-WISE-GAIA sample (pink) alongside the $N=13,724$ spectroscopically confirmed sources in the training set (blue) using a Mollweide projection in Galactic coordinates. While the training set is predominantly concentrated at high Galactic latitudes due to the footprint of major spectroscopic surveys like SDSS and DESI, our parent sample exhibits comprehensive coverage across the entire sky, including the Galactic plane ($|b| < 20^\circ$, shaded in light green). This visualization highlights that our machine-learning approach is strategically positioned to identify new candidates in regions where spectroscopic labels are currently sparse.

Although the training set is not strictly balanced, it provides broad coverage of photometric, astrometric, and X-ray parameter space for both quasars and stellar-type objects. This construction is designed to reduce overfitting and to enable the RF classifier described in Section \ref{sec:method} to robustly distinguish quasars from stellar contaminants in the low-latitude sky.

\section{Method}
\label{sec:method}

\subsection{The Random Forest Classifier}

\begin{figure*}[t]
    \centering
    \includegraphics[width=0.65\textwidth]{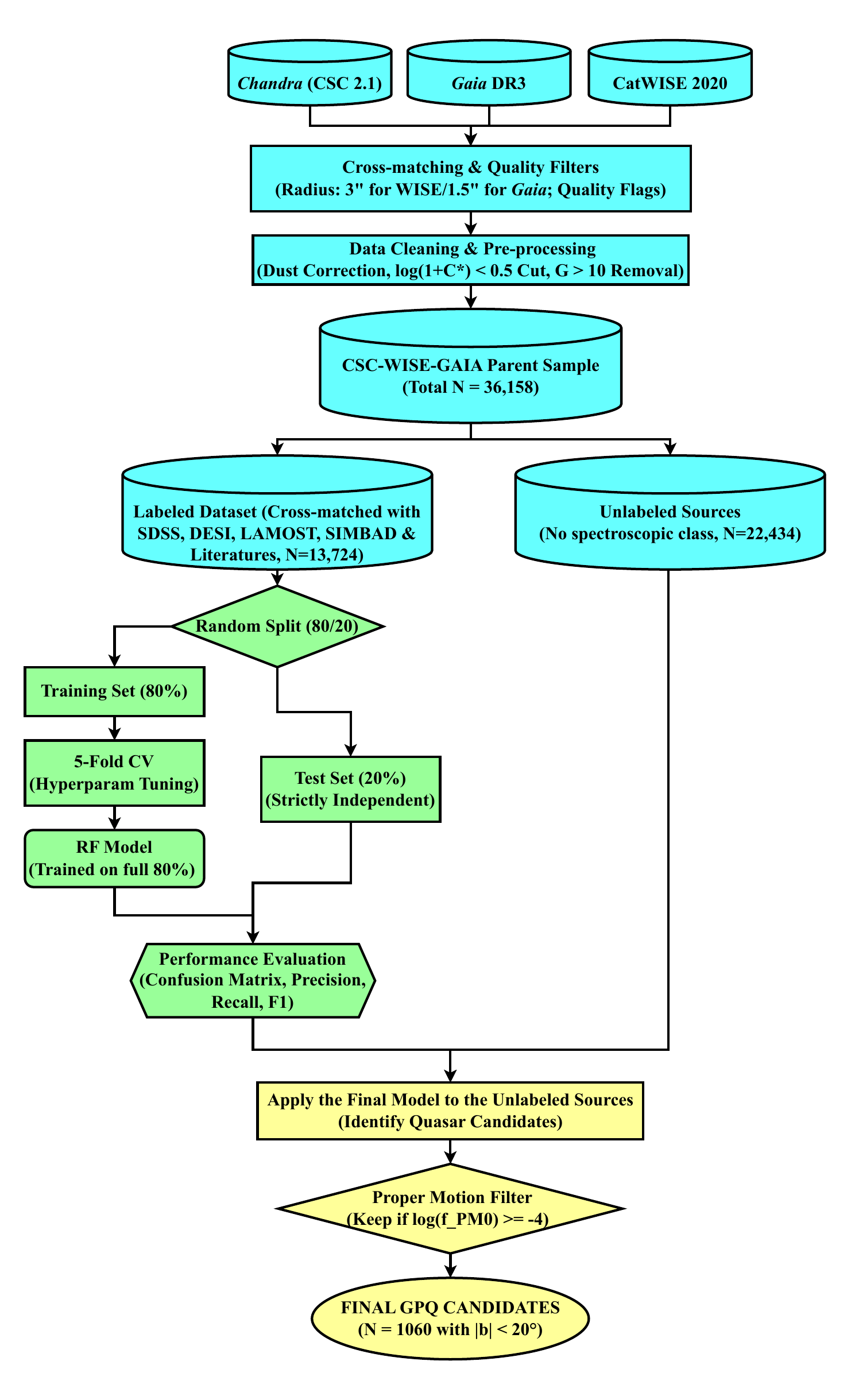}
    
    \caption{The flowchart of the machine learning pipeline used for GPQ candidate selection. The process begins with the construction of the multi-wavelength parent sample from \textit{Chandra}, \textit{Gaia}, and CatWISE. Crucially, to prevent data leakage, the labeled dataset ($N=13,724$) is split into a training set (80\%) and a strictly independent test set (20\%) before any hyperparameter tuning. The 5-fold cross-validation is performed exclusively within the training branch. The final Random Forest model is evaluated on the held-out test set before being applied to the unlabeled sources ($N=22,434$) to identify new candidates.}
    \label{fig:flowchart}
\end{figure*}

The overall workflow of our candidate selection pipeline is illustrated in Figure \ref{fig:flowchart}. We employ a RF classifier to distinguish quasars from stellar-type objects in the CSC--WISE--GAIA sample. RF is an ensemble learning algorithm based on decision trees and bootstrap aggregation, originally introduced by \citet{Ho1995} and later formalized by \citet{Breiman2001}. By combining multiple weak learners trained on bootstrapped subsets of the data and randomly selecting input features at each split, the RF algorithm effectively mitigates overfitting and performs robustly in high-dimensional parameter space.

The RF classifier is implemented using the \texttt{scikit-learn}\footnote{\url{https://scikit-learn.org/stable/index.html}} machine-learning library \citep{scikit-learn}. Each decision tree is trained independently, and the final classification is determined by a majority vote across all trees. For each source, the fraction of trees voting for the quasar class is interpreted as the quasar membership probability, providing a quantitative measure of classification confidence.

To separate quasars from stellar-type objects, we adopt a total of 31 input features that incorporate multi-wavelength photometric, astrometric, and X-ray information. These features include X-ray fluxes and hardness ratios from CSC~2.1, optical and mid-infrared magnitudes and colors from \textit{Gaia} DR3 and WISE, extinction-related quantities, and astrometric parameters such as proper motion and parallax significance. A complete list and description of the adopted features are provided in Table \ref{tab:features}.

\begin{table*}[t]
\caption{Features in This Work}
  \small
  \raggedright 
  \setlength{\tabcolsep}{15pt}
  \begin{tabular}{ll}
    \hline\hline
     \textbf{Feature}& \textbf{Feature description}\\
    \hline
                   gal\_l& Galactic Longitude\\
                   gal\_b& Galactic Latitude\\
                   $b$-csc& Flux in ACIS broad ($b$) band (0.5–7.0 keV)\\
                   $s$-csc& Flux in ACIS soft ($s$) band (0.5–1.2 keV)\\
                   $m$-csc& Flux in ACIS medium ($m$) band (1.2–2.0 keV)\\
                   $h$-csc& Flux in ACIS hard ($h$) band (2.0–7.0 keV)\\
                   $\mathrm{HR}_{hm}$& ACIS hard (2.0-7.0 keV) - medium (1.2-2.0 keV) energy band hardness ratio\\
                   $\mathrm{HR}_{hs}$& ACIS hard (2.0-7.0 keV) - soft (0.5-1.2 keV) energy band hardness ratio\\
                   $\mathrm{HR}_{ms}$& ACIS medium (1.2-2.0 keV) - soft (0.5-1.2 keV) energy band hardness ratio\\
                   $N_{\mathrm{H, Gal}}$& Galactic $N_H$ column density in direction of source\\
                   $\log(f_x)$& Logarithm of X-ray flux in 0.5-7.0 keV ($b$ band)\\
                   $\log(f_x/f_g)$& Logarithm of X-ray-to-optical flux ratio\\
    \hline
                   $G$& \textit{Gaia} DR3 $G$-band magnitude\\
                   $BP$& \textit{Gaia} DR3 $BP$-band magnitude\\
                   $RP$& \textit{Gaia} DR3 $RP$-band magnitude\\
                   $W1$& WISE $W1$ band magnitude\\
                   $W2$& WISE $W2$ band magnitude\\
                   $A_V$& Interstellar Extinction\\
                   $\log(f_g)$& Logarithm of $G$ band flux\\
                   $\log(f_{\mathrm{BP}})$& Logarithm of $BP$ band flux\\
                   $\log(f_{\mathrm{RP}})$& Logarithm of $RP$ band flux\\
                   PM& Total proper motion\\
                   PLXSIG& Parallax significance defined as $\left| \frac{\text{PARALLAX}}{\text{PARALLAX\_ERROR}} \right|$\\
                   PMSIG& Proper motion significance defined as $\sqrt{\left( \frac{\text{PMRA}}{\text{PMRA\_ERROR}} \right)^2+\left( \frac{\text{PMDEC}}{\text{PMDEC\_ERROR}} \right)^2}$\\
    \hline
                   $BP-RP$& $BP-RP$ color\\
                   $BP-G$& $BP-G$ color\\
                   $G-RP$& $G-RP$ color\\
                   $RP-W1$& $RP-W1$ color\\
                   $G-W1$& $G-W1$ color\\
                   $G-W2$& $G-W2$ color\\
                   $W1-W2$& $W1-W2$ color\\
    \hline
  \end{tabular}
  \label{tab:features}
\end{table*}

Missing values may arise from limited survey depth, nondetections, or incomplete coverage in specific bands.  To allow uniform processing of incomplete data, all missing entries are assigned a sentinel value of $-9999$. While large sentinel values can sometimes introduce artificial variance in distance-based models, Random Forest is naturally robust to them due to its tree-based architecture. The algorithm partitions the feature space using simple thresholds, effectively isolating sentinel values into independent branches without introducing artificial scaling issues. This approach preserves the physical information associated with missing data, allowing the model to handle such cases consistently during both training and prediction.

The RF model relies on several tunable hyperparameters whose optimization is essential for maximizing classification performance. To explicitly prevent data leakage, the complete labeled dataset ($N=13,724$) was first randomly partitioned into an 80\% training/validation set and a 20\% independent test set. To ensure an unbiased evaluation across the sky, this split was stratified by source type and regional category (i.e., within or outside the Galactic plane, defined as $|b| < 20^\circ$).

We focused on three key parameters: the number of trees in the ensemble (\texttt{n\_estimators}), the number of features considered at each split (\texttt{max\_features}), and the maximum depth of individual trees (\texttt{max\_depth}). Hyperparameter tuning was performed exclusively within the 80\% training set using a grid search with stratified five-fold cross-validation. The optimal configuration was determined to be $\texttt{max\_features} = \log_{2} 31$ and $\texttt{n\_estimators} = 215$, while no explicit upper limit on \texttt{max\_depth} was required based on the validation results.

\subsection{Classification Performance and Feature Importance}

Following parameter tuning, the final model was trained on the entire training set and evaluated on the untouched, strictly independent test set. Classification performance is assessed using standard metrics for binary classification, including accuracy, precision, recall, and $F1$-score, defined as:
\begin{equation}
\text{Accuracy} = \frac{\text{TP} + \text{TN}}{\text{TP} + \text{TN} + \text{FP} + \text{FN}}\,,
\end{equation}
\begin{equation}
\text{Precision} = \frac{\text{TP}}{\text{TP} + \text{FP}}\,,
\end{equation}
\begin{equation}
\text{Recall} = \frac{\text{TP}}{\text{TP} + \text{FN}}\,,
\end{equation}
\begin{equation}
F1 = 2 \times \frac{\text{Precision} \times \text{Recall}}{\text{Precision} + \text{Recall}}\,,
\end{equation}
where TP, TN, FP, and FN denote true positives, true negatives, false positives, and false negatives, respectively.

\begin{figure*}[t] 
  \includegraphics[width=\textwidth]{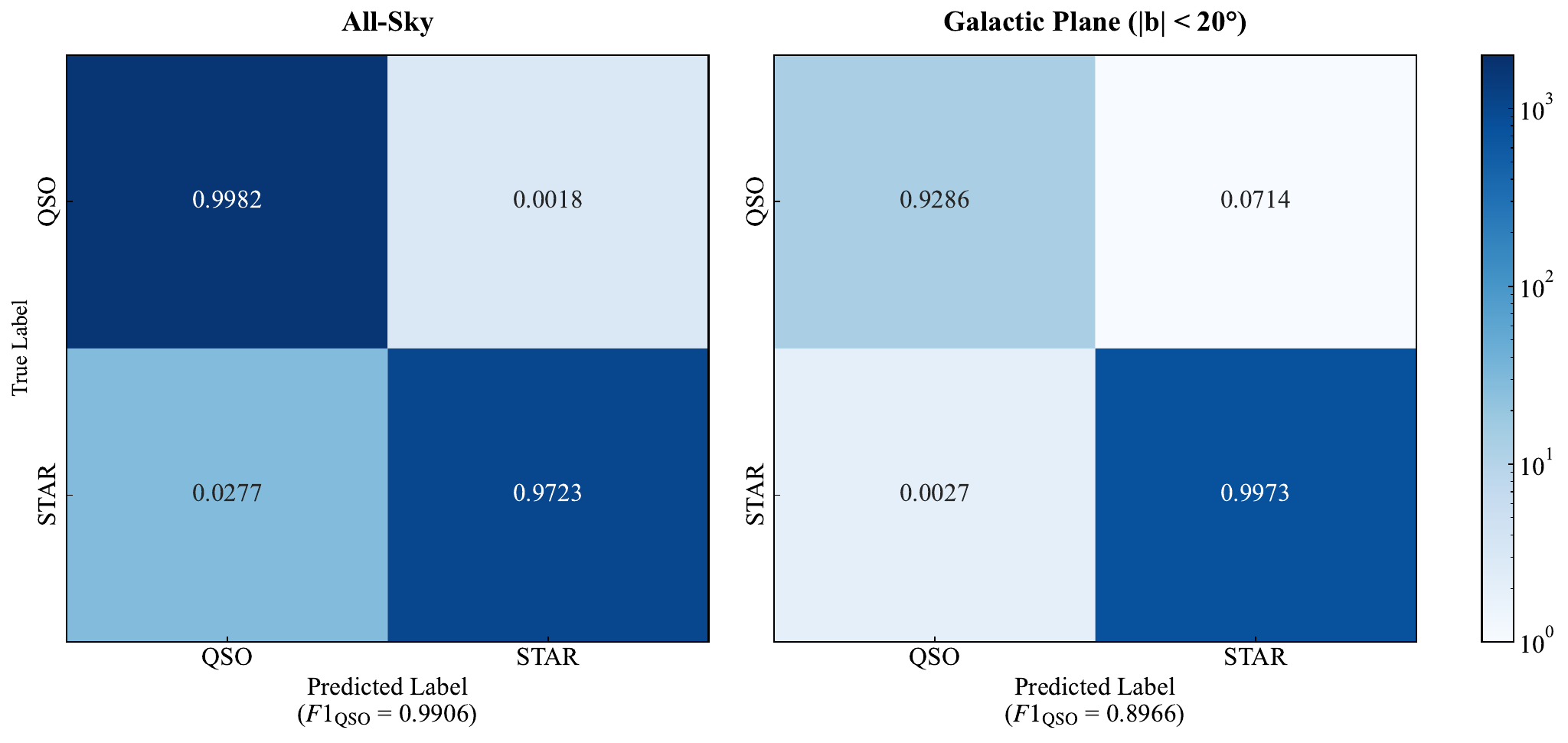}
  \caption{Normalized confusion matrices of the RF classifier evaluated on the independent test set for recall. The left panel shows the all-sky performance, while the right panel highlights the performance specifically within the Galactic plane ($|b| < 20^\circ$). The labels ``QSO'' and ``STAR'' represent quasars and stellar-type objects, respectively. The color scale uses a logarithmic normalization to better visualize the classification of minority classes.}
  \label{fig:confusion_matrix}
\end{figure*}

We first evaluate the performance of the trained RF classifier on the independent test set. As shown in the confusion matrices (Figure \ref{fig:confusion_matrix}), the classifier achieves a high overall accuracy of 98.83\%. However, considering the potential impact of the Galactic environment on the classification, we further assess the performance specifically at low Galactic latitudes ($|b| < 20^\circ$). While the all-sky F1-score for quasars is 99.06\%, it drops to 89.66\% in the Galactic plane region. These results are summarized in Table~\ref{tab:performance_comparison}. The slight decrease in performance at low Galactic latitudes is expected, given the effects of strong extinction and increased source confusion.

\begin{table}[b]
\caption{Comparison of Classification Performance between All-Sky and Galactic Plane Samples}
\setlength{\tabcolsep}{3pt}
\small
\label{tab:performance_comparison}
\begin{tabular*}{\columnwidth}{c|lccc}
\hline \hline
\textbf{Region} & \textbf{Class} & \textbf{Precision (\%)} & \textbf{Recall (\%)} & \textbf{F1 (\%)} \\
\hline
All-Sky & QSO & 98.32 & 99.82 & 99.06 \\
        & STAR & 99.71 & 97.23 & 98.45 \\
\hline
Galactic  & QSO & 86.67 & 92.86 & 89.66 \\
Plane     & STAR & 99.87 & 99.73 & 99.80 \\
\hline
\end{tabular*}

\vspace{10pt} 

\textbf{Note.}Metrics are calculated on the independent test set. The Galactic plane sample is defined as sources with $|b| < 20^\circ$. ``QSO'' and ``STAR'' represent quasars and stellar-type objects.
\end{table}

To further understand the limitations of our binary classifier and address the potential contamination from specific stellar types, we analyzed the subclass composition of the 29 false positives (i.e., stellar objects misclassified as quasars) in the independent test set.

Although our training set is dominated by normal stars ($N=3452$) and YSOs ($N=1601$), it also includes small but representative samples of accreting sources, including high-mass X-ray binaries (HMXBs, $N=27$), low-mass X-ray binaries (LMXBs, $N=22$), and cataclysmic variables (CVs, $N=23$).

Among the 29 false positives, the vast majority ($N=27$, 93.1\%) are normal stars. However, we also identified 2 X-ray binaries (1 HMXB and 1 LMXB) as contaminants. While XRBs constitute only $\sim0.4\%$ of the training set, they account for 6.9\% of the classification errors. This disproportionate representation confirms that accreting binaries, which share similar X-ray and optical properties with quasars, are indeed more challenging to distinguish.

In contrast, zero CVs were found among the false positives, indicating that our feature set effectively rejects these sources despite their accretion nature. Given the intrinsic rarity of XRBs compared to the vast population of field stars, their contribution to the total contamination remains critically low, justifying the effectiveness of our binary classification approach for the current sample size.

\begin{figure}[ht]
\centering
\includegraphics[width=\columnwidth]{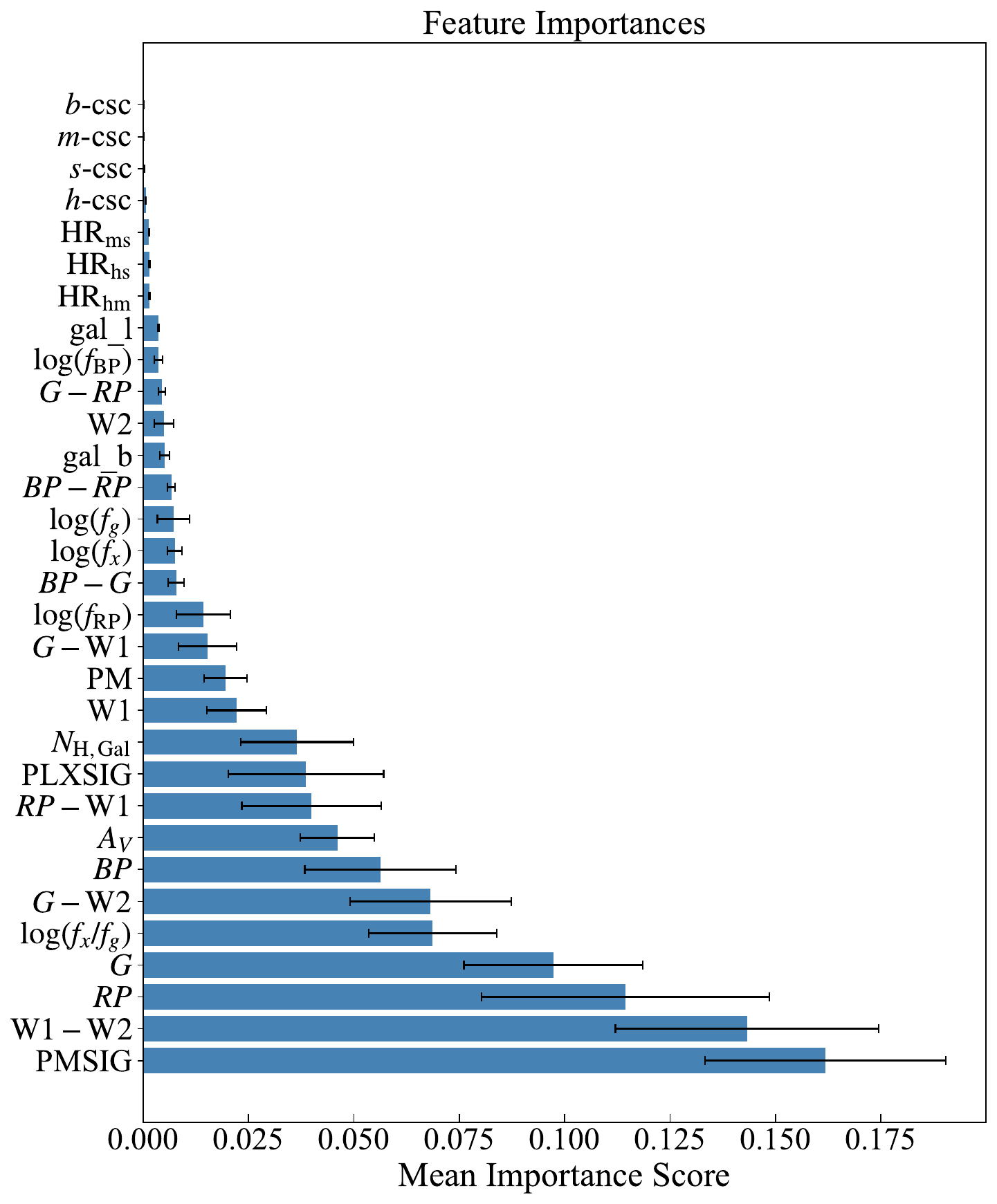}
\caption{Relative importance of the input features for the random forest classifier. Features are ranked based on their mean decrease in Gini impurity across 10 independent runs with different random seeds, with error bars representing the standard deviation. }
\label{fig:feature_importance}
\end{figure}

The feature importance analysis (Figure~\ref{fig:feature_importance}) reveals the physical drivers of our classification. The proper motion significance (PMSIG) emerges as the most critical feature, reflecting the fundamental astrometric difference between distant quasars, which are essentially stationary, and Galactic stars with detectable motions \citep[e.g.,][]{Gaia_2023a}. The mid-infrared color $W1 - W2$ also ranks highly, as it effectively captures the characteristic infrared excess produced by hot dust in the vicinity of active galactic nuclei, a feature that distinguishes quasars from most stellar populations. Furthermore, the importance of the $RP$ magnitude suggests that in heavily obscured regions like the Galactic plane, longer-wavelength optical photometry is vital for detecting background sources through interstellar extinction.

\section{Redshift Estimation for Quasar Candidates}\label{sec:redshift}

Applying the trained RF classifier to the 22,434 unclassified sources in the CSC--WISE--GAIA sample yields 7,681 quasar candidates. Within this sample, the 1,076 sources located at low Galactic latitudes ($|b| < 20^\circ$) are defined as Galactic Plane Quasar (GPQ) candidates (see Section \ref{subsec:5.1} for formal selection and further filtering). In this section, we use an RF-based regression model to estimate photometric redshifts for these candidates. Reliable redshift estimation is essential for characterizing the statistical properties of the candidate sample and for enabling subsequent population studies, particularly in regions where spectroscopic follow-up is incomplete.

The RF regression model is trained using spectroscopically confirmed quasars from the training set described in Section \ref{subsec:2.2}. To evaluate the performance of the regression, we adopt two commonly used metrics: the root mean square error (RMSE) and the coefficient of determination ($R^{2}$), defined as
\begin{equation}
\mathrm{RMSE} = \sqrt{\frac{1}{n} \sum_{i=1}^{n} (z_i - \hat{z}_i)^2},
\end{equation}
\begin{equation}
R^{2} = 1 - 
\frac{\sum_{i=1}^{n} (z_i - \hat{z}_i)^2}
{\sum_{i=1}^{n} (z_i - \bar{z})^2},
\end{equation}
where $z_i$ and $\hat{z}_i$ are the spectroscopic and predicted redshifts of the $i$th source, respectively, $\bar{z}$ is the mean spectroscopic redshift of the sample, and $n$ is the total number of sources. The RMSE quantifies the typical deviation between predicted and true redshifts, while $R^{2}$ measures the fraction of variance in the spectroscopic redshifts explained by the regression model.

Photometric redshift estimation based solely on \textit{Gaia} photometry is limited by the broad optical passbands and the lack of detailed spectral information. To improve regression performance, we incorporate additional optical photometry from the Pan-STARRS1 \citep[PS1;][]{PS1} survey, which provides five broadband filters ($g$, $r$, $i$, $z$, and $y$). For sources not covered by PS1, photometry from the NOIRLab Source Catalog DR2 \citep[NSC DR2;][]{NSC_DR2} is adopted as a substitute. These data enable the construction of more informative color indices that better trace redshift-dependent spectral features.

In total, we adopt 18 input features for the RF regression model. These include the optical, infrared, and astrometric color indices $g-r$, $r-i$, $i-z$, $z-y$, $z - W1$, $y - W1$, $W1 - W2$, $G_{\mathrm{BP}} - G_{\mathrm{RP}}$, $G_{\mathrm{BP}} - G$, $G - G_{\mathrm{RP}}$, $G_{\mathrm{RP}} - W1$, $G - W1$, and $G - W2$, as well as $\log(f_{x})$, $\log(f_{x}/f_g)$, $\log(1 + C^{*})$, $\log(1 + z_{\mathrm{low}})$, and $\log(1 + z_{\mathrm{up}})$. Here, $z_{\mathrm{low}}$ (\texttt{redshift\_qsoc\_lower}) and $z_{\mathrm{up}}$ (\texttt{redshift\_qsoc\_upper}) correspond to the 0.15866 and 0.84134 quantiles of the \textit{Gaia}-based redshift estimate $z_{\mathrm{Gaia}}$, representing the lower and upper confidence bounds, respectively.

\begin{figure}[t] 
  \includegraphics[width=\columnwidth]{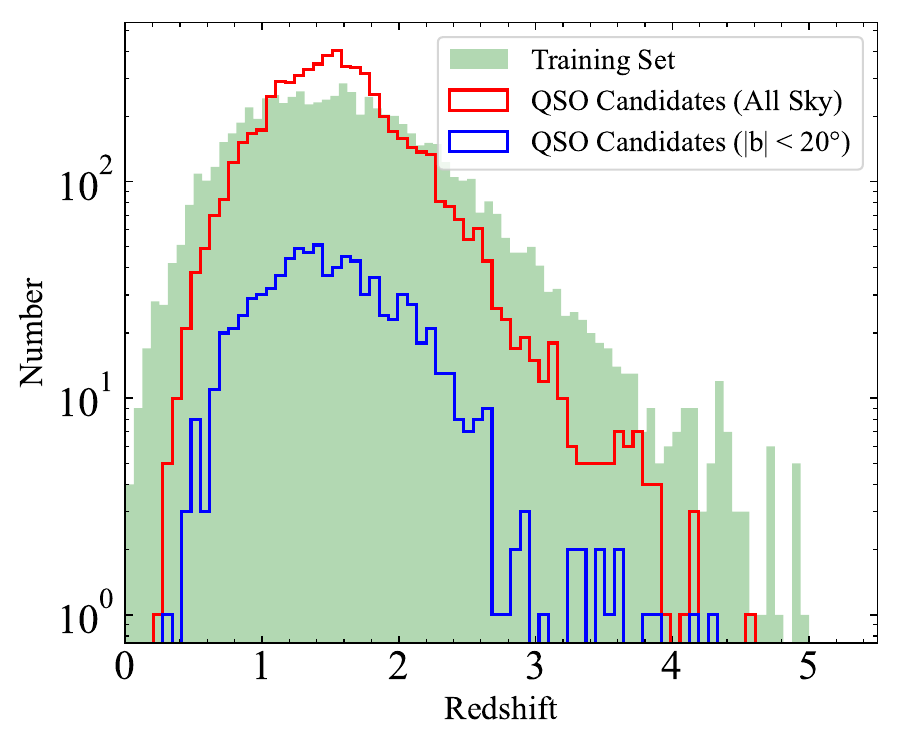}
  \caption{Distribution of redshifts. The green filled histogram displays the spectroscopic redshift distribution of the training/test sample ($N=$7,332) used to build the regression model.} The red open histogram shows the RF-based photometric redshifts for the full-sky candidate sample, and the blue open histogram shows the candidates at low Galactic latitude ($|b|<20^\circ$). The ordinate shows source counts (logarithmic scale).
  \label{fig:redshift}
\end{figure}

The final regression sample, constructed from the intersection of our labeled training set and the PS1/NSC photometric catalogs, consists of 7,332 spectroscopically confirmed quasars. The redshift distribution of this spectroscopic sample is shown as the green filled histogram in Figure~\ref{fig:redshift}, covering a wide range of $0 < z < 5$ which effectively spans the expected redshift space of our candidates.

\begin{figure}[t]
\centering
\includegraphics[width=\columnwidth]{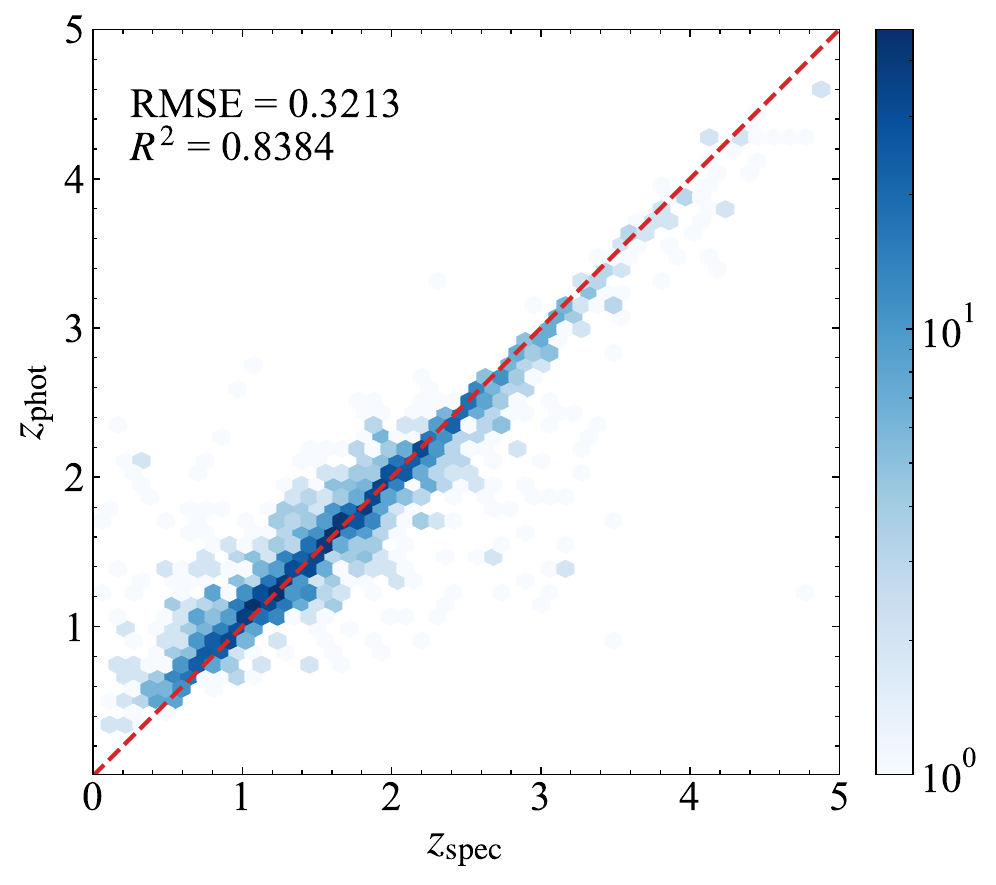} 
\caption{Quantitative evaluation of the regression model for photometric redshift estimation. The hexagonal binning plot displays the density of predicted redshifts ($z_{\mathrm{phot}}$) versus spectroscopic redshifts ($z_{\mathrm{spec}}$) for the independent test set. The logarithmic color scale indicates the source count per bin.}
\label{fig:redshift_eval}
\end{figure}

We performed a quantitative evaluation of the regression model using the independent test set. The results are summarized in Figure~\ref{fig:redshift_eval}. The model yields an RMSE of 0.3213 and a coefficient of determination ($R^2$) of 0.8384. These metrics, combined with the strong concentration of sources along the identity line in the density plot, demonstrate that our photometric redshift estimates are robust and free from significant latitude-dependent systematic biases.

The trained RF regression model is applied to all quasar candidates to derive photometric redshift estimates. Figure \ref{fig:redshift} shows the resulting redshift distributions for the all-sky quasar candidate sample and for the subset of GPQ candidates. The redshift distribution of the GPQs is broadly consistent with that of the all-sky sample, indicating that the RF regression model does not introduce significant redshift-dependent biases in the low-latitude regime. These photometric redshifts provide a useful statistical characterization of the quasar candidate population and serve as a basis for future spectroscopic follow-up observations.

\section{Results}\label{sec:result}

\subsection{GPQ candidate selection and additional filtering}\label{subsec:5.1}

\begin{figure}[t]
  \centering
  \includegraphics[width=\columnwidth]{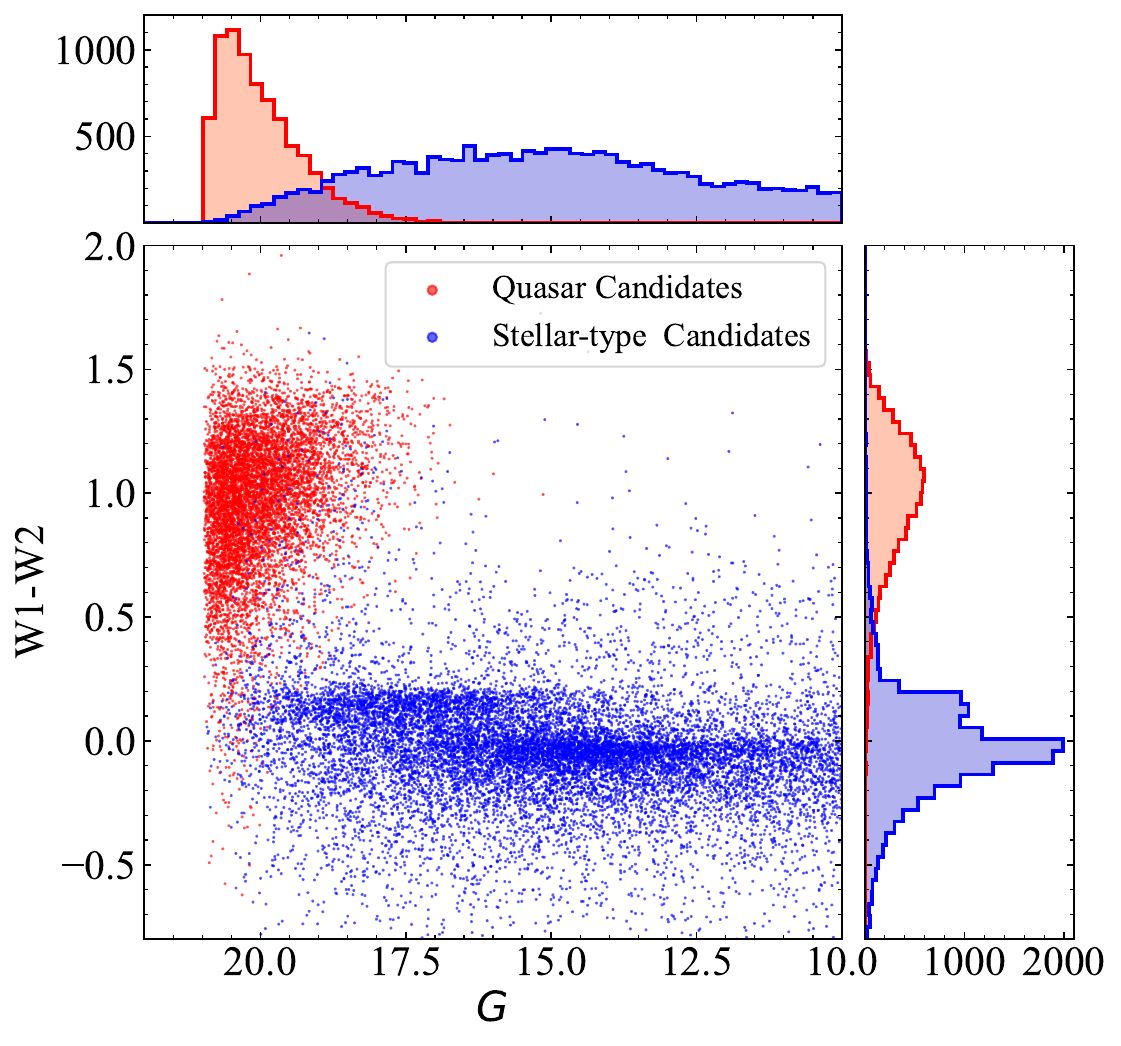}
 \caption{Distribution of the all-sky candidates in the $G$ versus $W1 - W2$ diagram as classified by the RF model. Red and blue points denote quasar and stellar-type candidates, respectively; the top and right panels show the corresponding marginal distributions in $G$ and $W1 - W2$. The $G$ magnitudes shown on the horizontal axis are corrected for Galactic extinction. We apply a uniform magnitude filter to all candidates, excluding sources with observed $G>21$~mag (prior to extinction correction) and removing very bright objects with $G<10$~mag.}
  \label{fig:gmag-w1w2}
\end{figure}

We apply the trained RF classifier to the 22,434 previously unclassified sources in the CSC-WISE-GAIA sample, i.e., sources without reliable spectroscopic classifications. Of these, 7681 sources are classified as quasar candidates and 14,753 as stars. Figure \ref{fig:gmag-w1w2} presents the distribution of the resulting quasar and stellar candidates in the $G$ versus $W1 - W2$ color--magnitude diagram.

Among the previously unclassified sources, 11,751 are located within the Galactic plane ($|b| < 20^{\circ}$). Of these, \textbf{1076} sources are classified as quasar candidates, while the remaining sources are classified as stellar-type objects, reflecting the strong dominance of Galactic stellar populations at low Galactic latitudes.

To further suppress stellar contamination, we apply an additional selection based on \textit{Gaia} proper motions. We adopt the probabilistic zero--proper-motion criterion introduced by \citet{Fu_2021,Fu_2024,Fu_2025}, which explicitly accounts for measurement uncertainties. The probability density of zero proper motion, $f_{\mathrm{PM}0}$, is defined as

\begin{equation}
\begin{split}
f_{\mathrm{PM}0} &= \frac{1}{2\pi\sigma_x\sigma_y\sqrt{1-\rho^2}} \times \\
&\quad \exp\left[
-\frac{1}{2(1-\rho^2)}
\left(
\frac{x^2}{\sigma_x^2}
-\frac{2\rho xy}{\sigma_x\sigma_y}
+\frac{y^2}{\sigma_y^2}
\right)\right]\,,
\end{split}
\label{eq:fpm0}
\end{equation}

\noindent where $x$ is the proper motion in R.A. (\texttt{pmra}), $y$ is the proper motion in Decl. (\texttt{pmdec}), $\sigma_x$ and $\sigma_y$ are the corresponding uncertainties, and $\rho$ is the correlation coefficient between
$x$ and $y$ (\texttt{pmra\_pmdec\_corr}). For a given uncertainty level, sources with smaller proper motions yield higher values of $f_{\mathrm{PM}0}$ by construction.

\begin{figure}[t]
  \centering
  \includegraphics[width=\columnwidth]{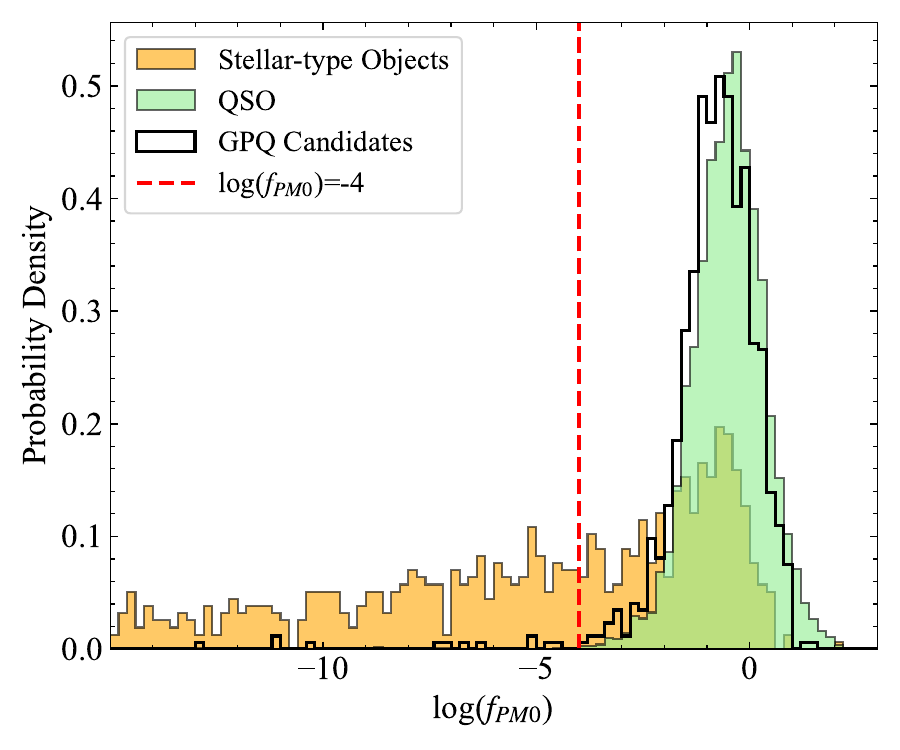}
  \caption{Distributions of $\log(f_{\mathrm{PM}0})$, the probability density at zero proper motion derived from \textit{Gaia} astrometry (see Eq.~\ref{eq:fpm0}), for spectroscopically confirmed quasars (green), spectroscopically confirmed stellar-type objects (yellow), and the GPQ candidates (white). The vertical dashed line marks the adopted threshold $\log(f_{\mathrm{PM}0})=-4$ used to suppress stellar contaminants; note that $f_{\mathrm{PM}0}$ is a probability density (not a probability) and therefore can exceed unity.}
  \label{fig:logfPM0_dist}
\end{figure}

Figure \ref{fig:logfPM0_dist} shows the distributions of $\log(f_{\mathrm{PM}0})$ for spectroscopically confirmed quasars, spectroscopically confirmed stellar-type objects, and the GPQ candidates. We adopt a conservative threshold of $\log(f_{\mathrm{PM}0}) \geq -4$, which efficiently removes stellar contaminants while retaining the majority of quasars.

\begin{figure*}[t] 
  \includegraphics[width=\textwidth]{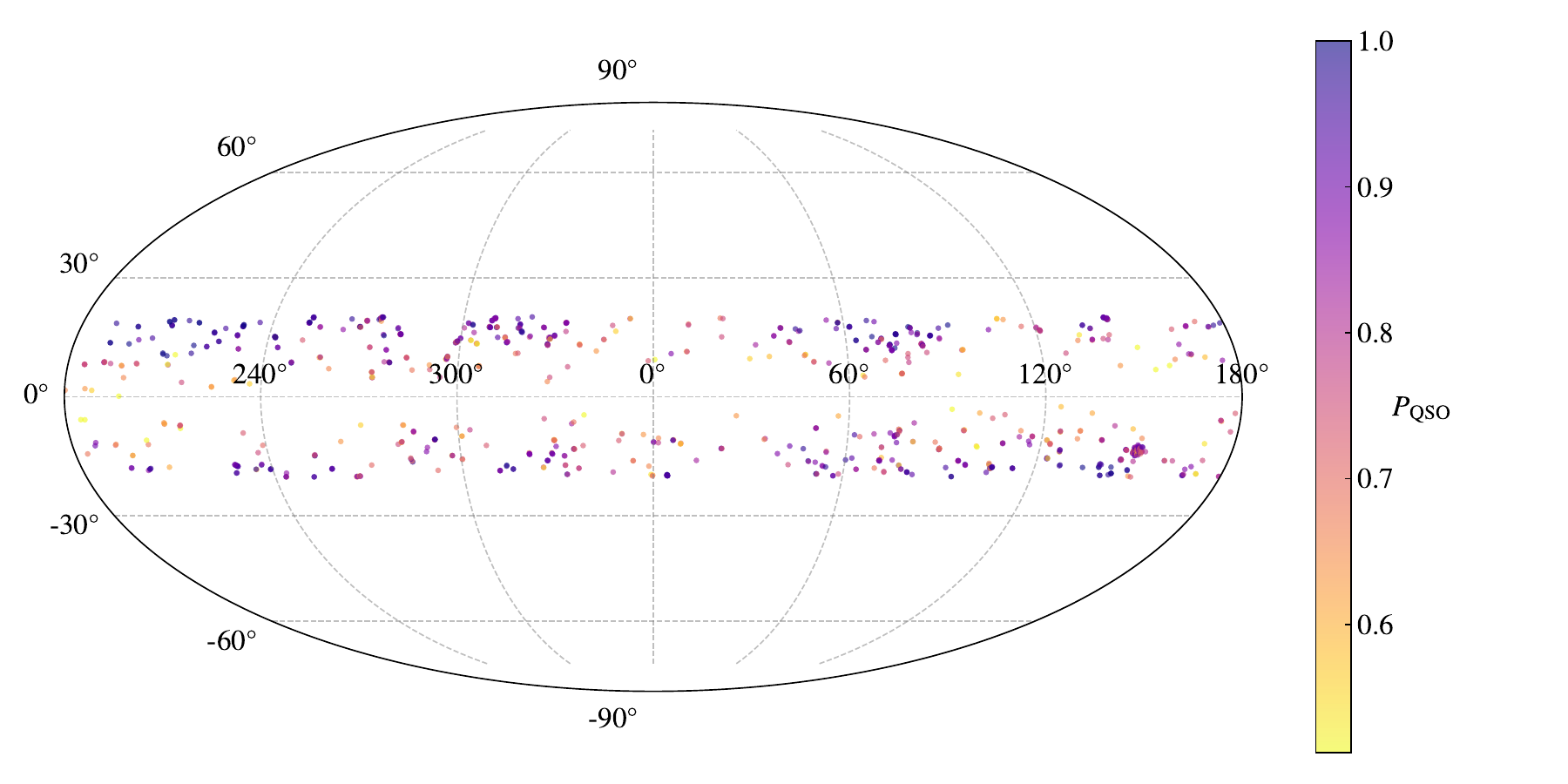}
  \caption{Spatial distribution of GPQ candidates ($|b|<20^\circ$) in Galactic coordinates (Mollweide projection). Points are color-coded by the RF-derived quasar membership probability, $P_{\mathrm{QSO}}$ (color bar).}
  \label{fig:source_map}
\end{figure*}

After the proper-motion filtering, the final quasar candidate sample contains 7570 sources across the entire sky. Among these, 1060 sources are located behind the Galactic plane ($|b| < 20^{\circ}$) and are identified as GPQ candidates. Of the GPQs, 551 sources have quasar membership probabilities greater than 0.8 and are therefore considered high-confidence GPQ candidates. The spatial distribution of the GPQ candidates in Galactic coordinates is shown in Figure \ref{fig:source_map}. A subset of these high-confidence GPQ candidates is presented in Table~\ref{tab:candidate_sources}.

As mentioned in Section \ref{subsec:2.1}, we evaluated the impact of source confusion in the Galactic plane using the probabilistic algorithm \texttt{nway}. Applying strict probability cuts with \texttt{nway} yielded 980 GPQ candidates, representing a relatively small reduction from the 1,060 candidates obtained via our nearest-neighbor approach. More importantly, we found that the strict \texttt{nway} procedure completely rejected two of our high-confidence candidates that were recently spectroscopically confirmed as quasars via the Next Generation Palomar Spectrograph (NGPS) on the Hale Telescope (see Section \ref{sec:spec_followup}). This observation directly confirms that while probabilistic matching filters out ambiguous blended sources, it penalizes completeness. Because our primary goal is to identify more quasar candidates in the Galactic plane, we retain the nearest-neighbor matching results.

\subsection{Properties of GPQ Candidates}
\label{subsec:gpqprop}

\begin{table*}[t]
\caption{Selected GPQ Candidates by the RF Classifier. An excerpt of the GPQ candidate catalog is shown for guidance. The full machine-readable table is available at \url{https://github.com/jeepsmoker288/GPQs-and-all-sky-quasar-candidates}.}
\raggedright
\setlength{\tabcolsep}{9pt} 
\begin{tabular*}{\textwidth}{lrrcccccc} 
\hline\hline
\text{Name} & \text{R.A.} & \text{Decl.} & $W1$ & $W2$ & $G$ & $\log(f_x)$ & $P_{\mathrm{QSO}}$ & $z_{\mathrm{phot}}$\\ 
\hline
2CXO J011808.4+453857 & 19.5354 & 45.6492 & 15.244 & 14.510 & 20.2034 & -13.5972 & 0.9756 & 1.1070\\
2CXO J022142.3+421947 & 35.4264 & 42.3297 & 16.116 & 15.646 & 20.0635 & -14.1680 & 0.9756 & 3.4535\\
2CXO J031404.4+403900 & 48.5184 & 40.6500 & 15.697 & 14.590 & 19.3212 & -13.1726 & 0.9024 & 1.1544\\
2CXO J031711.4+421654 & 49.2975 & 42.2819 & 16.092 & 15.515 & 20.5539 & -13.7589 & 0.9512 & 1.1133\\
2CXO J031712.7+405301 & 49.3031 & 40.8838 & 14.768 & 13.349 & 19.2251 & -12.7701 & 0.8537 & 0.7809\\
2CXO J040239.4+260839 & 60.6643 & 26.1442 & 16.309 & 15.383 & 20.3126 & -13.6567 & 0.9512 & 1.2689\\
2CXO J061640.3-215410 & 94.1681 & -21.9029 & 16.929 & 16.060 & 20.5411 & -14.0410 & 0.9268 & 2.0291\\
2CXO J062659.7-352937 & 96.7488 & -35.4936 & 15.620 & 14.678 & 20.3614 & -13.8247 & 0.9024 & 1.6355\\
2CXO J075142.4-014522 & 117.9268 & -1.7564 & 15.725 & 15.314 & 20.7567 & -13.4226 & 0.9268 & 1.8400\\
2CXO J075255.9+122712 & 118.2330 & 12.4535 & 15.733 & 15.107 & 18.8968 & -13.5895 & 1.0000 & 3.5815\\
2CXO J081813.4-073306 & 124.5558 & -7.5520 & 16.557 & 16.032 & 20.0714 & -13.7727 & 0.9268 & 1.7899\\
2CXO J082705.8-070859 & 126.7744 & -7.1499 & 15.624 & 15.286 & 20.1933 & -14.1581 & 0.9024 & 3.2615\\
2CXO J082732.9-070831 & 126.8873 & -7.1420 & 16.023 & 14.801 & 19.9085 & -14.0502 & 0.9756 & 2.0211\\
2CXO J084138.9-173123 & 130.4121 & -17.5232 & 16.396 & 15.849 & 20.6588 & -14.0028 & 0.9024 & 1.6542\\
2CXO J100023.3-302411 & 150.0970 & -30.4030 & 15.895 & 14.910 & 20.3972 & -13.7827 & 0.9756 & 1.0419\\
2CXO J102512.1-472309 & 156.3008 & -47.3860 & 15.727 & 14.932 & 20.3067 & -13.4771 & 0.9756 & 0.9385\\
2CXO J104151.1-704759 & 160.4631 & -70.7999 & 15.621 & 14.776 & 20.1958 & -13.2347 & 0.9756 & 1.2306\\
\hline
\end{tabular*}

\vspace{10pt} 

\textbf{Note.} All $G$ band magnitudes have been corrected for Galactic dust extinction using the \citet{Planck} dust map. $\log(f_x)$ is the logarithm of the X-ray flux in 0.5--7.0 keV ($b$ band). $P_{\mathrm{QSO}}$ is the classification probability of QSOs. $z_{\mathrm{phot}}$ is the photometric redshift.
\label{tab:candidate_sources}
\end{table*}

\begin{figure}[t] 
  \includegraphics[width=\columnwidth]{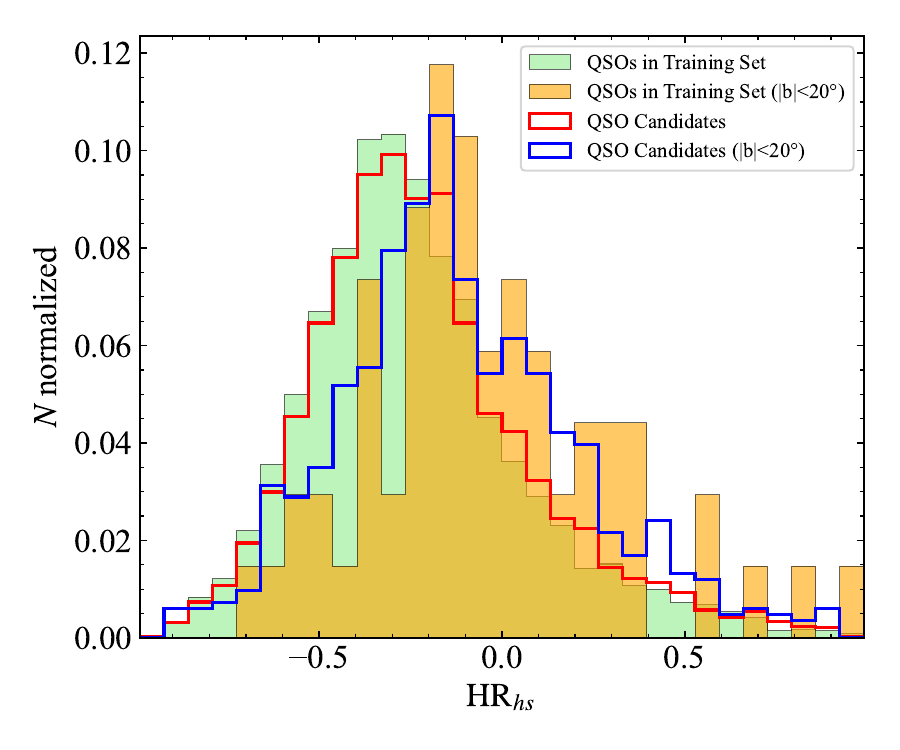}
  \caption{Normalized histograms of the $\mathrm{HR}_{hs}$ (hard – soft energy band hardness ratio, see Table \ref{tab:features}) for quasars. Filled histograms represent all-sky/Galactic plane ($|b|<20^\circ$) training set samples, and step outlines represent selected all-sky/Galactic plane candidates.}
  \label{fig:hard_hs_dist}
\end{figure}

The GPQ candidates identified in this work exhibit distinct X-ray and multi-wavelength properties compared with the all-sky quasar candidate sample. In particular, GPQs show systematically harder X-ray spectra. Figure \ref{fig:hard_hs_dist} compares the distributions of the hard--soft energy band hardness ratio ($\mathrm{HR}_{hs}$) for quasars in the spectroscopic training set and for the quasar candidates, separated into all-sky and low Galactic latitude ($|b| < 20^{\circ}$) subsamples. In both the training set and the candidate sample, sources located at $|b| < 20^{\circ}$ display higher hardness ratios than all-sky sources.

This systematic hardening is not indicative of intrinsic differences in quasar accretion properties, but instead reflects the effects of strong Galactic absorption along low-latitude sightlines. Soft X-ray photons are preferentially absorbed by interstellar gas and dust in the Galactic plane, whereas higher-energy photons are less affected. The close agreement between the hardness-ratio distributions of the GPQ candidates and those of spectroscopically confirmed GPQs in the training set indicates that the RF classifier effectively captures this absorption-driven spectral signature.

\begin{figure*}[t]
  \centering
  \includegraphics[width=0.9\textwidth]{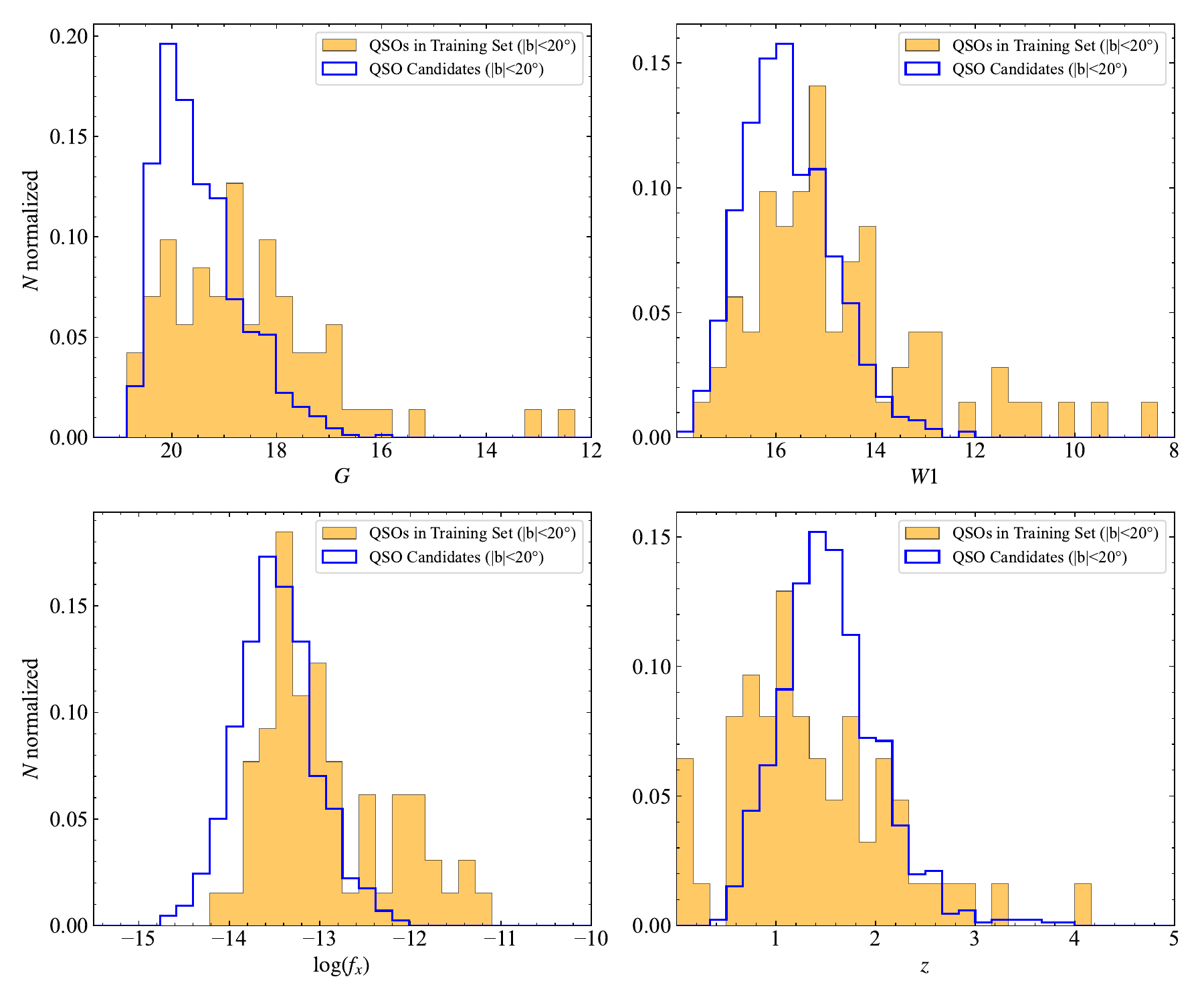}
  \caption{Normalized histograms of $G$, $W1$, $\log(f_x)$ (logarithm of X-ray flux in 0.5--7.0 keV), and redshift (here, the redshifts of the QSO candidates are estimated photometrically via RF) for GPQs and GPQ candidates. Filled histograms represent training set samples, and step outlines represent selected candidates.}
  \label{fig:4plot}
\end{figure*}

Figure \ref{fig:4plot} compares the distributions of optical ($G$), mid-infrared ($W1$), and X-ray fluxes, as well as redshift (spectroscopic or RF-based photometric redshift), between the training-set quasars and the GPQ candidates. In all wavelength regimes, the GPQ candidates are systematically fainter than the quasars in the training set. The $G$ and $W1$ magnitude distributions are shifted toward higher values, and the X-ray flux distribution peaks at lower fluxes.

These trends indicate that the present selection method is sensitive to quasars that lie beyond the effective depth of existing spectroscopic samples. The higher optical and infrared magnitudes of the GPQ candidates are consistent with significant dust extinction in the Galactic plane, while their lower X-ray fluxes suggest that this work probes a previously underexplored population of faint or heavily obscured quasars behind the Galactic disk.

We further compare the GPQ candidates identified in this work with the unified all-sky quasar candidate catalog CatGlobe \citep{Fu_2025}. Of the 1060 GPQ candidates, 600 have counterparts in CatGlobe, while 460 do not. The unmatched candidates show no significant differences from the matched subset in their color--magnitude distributions, X-ray fluxes, or redshift distributions, indicating that they likely belong to the same underlying quasar population. The presence of these additional candidates highlights the importance of incorporating X-ray information in quasar selection, particularly for uncovering quasars missed by purely optical and infrared approaches in the heavily obscured regions of the Galactic plane.

\begin{figure*}[t]
  \centering
  \includegraphics[width=0.95\textwidth]{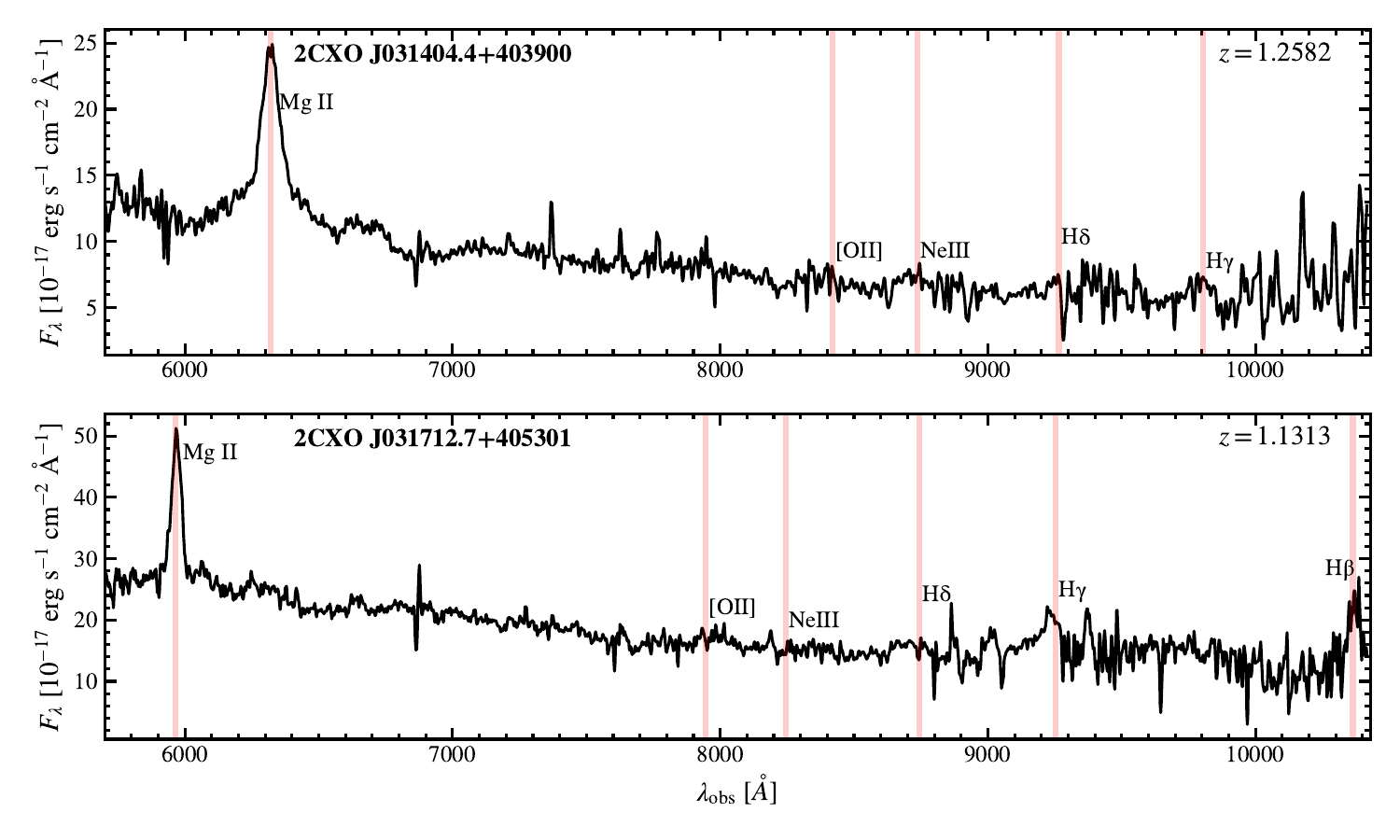}
  \caption{Spectra of the two identified GPQs from our pilot follow-up program with the Next Generation Palomar Spectrograph (NGPS) on the Hale Telescope. We show flux- and wavelength-calibrated optical spectra of \object{2CXO J031404.4+403900} (upper panel) and \object{2CXO J031712.7+405301} (lower panel). The redshift inferred from identified emission lines is given in the upper-right corner of each panel. Strong emission features are labeled, and the spectra are displayed in the observed frame.}
  \label{fig:spectra}
\end{figure*}

\subsection{Spectroscopic Follow-up of GPQ Candidates}\label{sec:spec_followup}

The most critical next step is spectroscopic confirmation of the candidate GPQs. This will definitively validate the RF method's accuracy, measure precise redshifts, and allow for detailed studies of the physical properties (e.g., black hole masses, accretion rates) of these previously hidden quasars. To provide an initial validation of our GPQ candidate selection, we obtained pilot optical spectroscopy for two high-confidence GPQ candidates, \object{2CXO J031404.4+403900} and \object{2CXO J031712.7+405301}. The observations were carried out on 2025 December 14 (UT) with the Palomar 200-inch (P200) Hale telescope using the Next Generation Palomar Spectrograph \citep[NGPS;][]{Jiang_2018}. NGPS records simultaneous spectra in two channels, covering approximately 580--780\,nm (R channel) and 760--1040\,nm (I channel). We used a $1\farcs5$ slit for both targets. 

The NGPS data were reduced with the \textsc{PypeIt}\footnote{\url{https://pypeit.readthedocs.io/en/release/index.html}} spectroscopic reduction pipeline \citep{Prochaska2020}, following standard long-slit procedures. Briefly, the two-dimensional frames were bias/overscan corrected and flat-fielded, after which the object trace was determined in each channel. Sky/background was estimated from source-free regions adjacent to the trace and subtracted, and one-dimensional spectra were extracted with a fixed-width aperture centered on the trace. Wavelength solutions were derived from arc-lamp exposures obtained in the standard NGPS calibration sequence. The R- and I-channel spectra were then placed on a consistent relative scale and combined by normalizing in their overlap region to produce a continuous spectrum for each target. Standard-star observations from the same run were used for relative flux calibration and to mitigate telluric absorption features.

 In Figure~\ref{fig:spectra}, we present the spectra of the two identified GPQs. Redshifts were measured from identified broad emission lines, using multiple features where available. For \object{2CXO J031404.4+403900}, we detect prominent Mg~\textsc{ii} together with [O~\textsc{ii}] and [Ne~\textsc{iii}], as well as Balmer emission (H$\delta$, H$\gamma$, and H$\beta$), yielding a spectroscopic redshift of $z_{\mathrm{spec}}=1.2582$. This measurement is in good agreement with our photometric estimate ($z_{\mathrm{phot}}=1.1544$). For \object{2CXO J031712.7+405301}, Mg~\textsc{ii}, [O~\textsc{ii}], and [Ne~\textsc{iii}] are detected along with Balmer emission, determining $z_{\mathrm{spec}}=1.1313$. The corresponding photometric redshift is $z_{\mathrm{phot}}=0.7809$. While this deviates by $\Delta z = 0.3504$, it remains consistent with the global RMSE ($0.3213$) of our regression model, representing a typical $\sim 1\sigma$ scatter. These confirmations provide a proof of concept that our X-ray plus multi-wavelength RF framework can recover genuine quasars behind the Galactic plane and motivate more extensive spectroscopic follow-up of the full GPQ candidate sample.

\section{Summary and Conclusions}\label{sec:summary}

In this work, we present a systematic search for quasars located behind the Galactic plane using X-ray–selected sources from the \emph{Chandra} Source Catalog (CSC~2.1), combined with optical and mid-infrared data from \textit{Gaia} DR3 and CatWISE2020. By exploiting the high angular resolution of \emph{Chandra} and the complementary diagnostic power of multi-wavelength information, we address the long-standing incompleteness of quasar samples at low Galactic latitudes.

We construct a high-quality CSC--WISE--GAIA sample through a series of photometric, astrometric, and morphological quality cuts designed to mitigate source confusion and contamination in crowded fields. Using spectroscopically confirmed quasars and stellar-type objects compiled from DESI, SDSS, LAMOST, and the literature, we train a RF classifier to distinguish quasars from Galactic sources. To further suppress stellar contamination, we apply a probabilistic zero–proper-motion criterion based on \textit{Gaia} astrometry. In addition, we estimate photometric redshifts for the quasar candidates using a RF–based regression model incorporating optical, infrared, and X-ray features.

Applying this framework to previously unclassified CSC sources, we identify a total of 7570 quasar candidates across the sky. Among these, 1060 sources are located behind the Galactic plane ($|b| < 20^{\circ}$) and are identified as GPQ candidates, with 551 classified as high-confidence candidates based on their quasar membership probabilities ($P_{\mathrm{QSO}} > 0.8$). Pilot optical spectroscopy confirms two high-confidence GPQ candidates, \object{2CXO~J031404.4+403900} ($z_{\mathrm{spec}}=1.2582$) and \object{2CXO~J031712.7+405301} ($z_{\mathrm{spec}}=1.1313$), providing direct validation of our selection approach. Motivated by these initial confirmations, we will carry out more extensive spectroscopic follow-up of the full GPQ candidate sample.

The GPQ candidates exhibit systematically harder X-ray spectra than their high-latitude counterparts, a trend that is consistently observed in both the training sample and the newly identified candidates. This behavior is naturally explained by strong Galactic absorption along low-latitude sightlines, which preferentially attenuates soft X-ray photons. In addition, the GPQ candidates are systematically fainter in the optical, infrared, and X-ray bands compared with spectroscopically confirmed quasars, indicating that this work probes a population of faint or heavily obscured quasars that is largely inaccessible to purely optical or infrared selection methods.

Comparison with the unified all-sky quasar candidate catalog CatGlobe shows that while 600 of our GPQ candidates have counterparts in existing catalogs, a substantial fraction (460 sources) are newly identified. These additional candidates exhibit properties consistent with the broader GPQ population, underscoring the importance of incorporating X-ray information to achieve a more complete census of quasars in the Galactic plane.

Overall, this study demonstrates that X-ray–based selection with \emph{Chandra}, combined with multi-wavelength data and modern classification techniques, provides an effective pathway for identifying quasars in the most heavily obscured regions of the sky. The resulting GPQ candidate sample represents a valuable resource for future spectroscopic follow-up, for improving the uniformity of all-sky quasar catalogs, and for applications ranging from Galactic absorption studies to the construction of astrometric reference frames in the Galactic disk. Future extensions of this work to larger X-ray datasets and upcoming facilities will further enhance our ability to uncover quasars hidden behind the Milky Way. The significant increase of the GPQ sample will help improve the reference frame for astrometry by using GPQs as background references and probe the Milky Way interstellar and circumgalactic media with the absorption features of GPQs.

\begin{acknowledgments}
We are grateful to the reviewer for his valuable comments and constructive suggestions, which have significantly improved the quality of our manuscript. We acknowledge the supports from National Natural Science Foundation of China (NSFC; grant Nos. 12133001, 12573110, 12273076), and National Key R\&D Program of China (2025YFA1614101). We acknowledge the support of Shenzhen Science and Technology program (JCYJ20230807113910021) and the Natural Science Foundation of Top Talent of SZTU (GDRC202208)

This research made use of data from the \emph{Chandra} X-ray Observatory, in particular the Chandra Source Catalog (\href{https://cxc.cfa.harvard.edu/csc/}{CSC website}) provided by the Chandra X-ray Center (CXC). It also made use of data from the European Space Agency (ESA) mission \emph{Gaia} (\href{https://www.cosmos.esa.int/web/gaia}{Gaia website}), processed by the \emph{Gaia} Data Processing and Analysis Consortium (\href{https://www.cosmos.esa.int/web/gaia/dpac/consortium}{DPAC consortium}). We used mid-infrared photometry from WISE/NEOWISE via the CatWISE2020 catalog and made use of the NASA/IPAC Infrared Science Archive (IRSA), which is funded by NASA and operated by the California Institute of Technology. Optical photometry was taken from the Pan-STARRS1 Surveys (\href{https://outerspace.stsci.edu/spaces/PANSTARRS/overview}{PS1}) and, where PS1 coverage is unavailable, from the NOIRLab Source Catalog DR2 (\href{https://datalab.noirlab.edu/data/nsc}{NSC website}). We used dust reddening information based on \emph{Planck}. We made use of public spectroscopic classifications from DESI DR1, SDSS DR18, and LAMOST; LAMOST is a National Major Scientific Project built by the Chinese Academy of Sciences, funded by the National Development and Reform Commission, and operated and managed by the National Astronomical Observatories, Chinese Academy of Sciences. This research has made use of the SIMBAD database, operated at CDS, Strasbourg, France, and services provided by Astro Data Lab, part of the Community Science and Data Center Program of NSF NOIRLab, operated by AURA under a cooperative agreement with the U.S. National Science Foundation.

Spectroscopic observations were obtained with the Palomar 200-inch Hale Telescope at Palomar Observatory using the Next Generation Palomar Spectrograph (\href{https://www.astro.caltech.edu/ngps}{NGPS}). We thank the Palomar Observatory staff for their support during the observations.

For DESI and SDSS, we acknowledge the respective collaborations and data release teams, their funding sources, and the observatory sites. In particular, we recognize the cultural significance of I'oligam Du'ag (Kitt Peak) to the Tohono O'odham Nation.

This work made use of TOPCAT and \texttt{scikit-learn} for catalog cross-matching and machine-learning analyses, and \textsc{PypeIt} for spectroscopic data reduction.

\end{acknowledgments}

\bibliography{references}{} 
\bibliographystyle{aasjournalv7}

\end{document}